\documentstyle[aps,preprint]{revtex}

\newcommand{\amu}{\alpha(\mu)}
\newcommand{\ame}{\alpha(m_e)}
\newcommand{\ammu}{\alpha(m_\mu)}
\newcommand{\amupms}{\alpha(\mu_\textrm{\tiny PMS})}
\newcommand{\amublm}{\alpha(\mu_\textrm{\tiny BLM})}

\tightenlines

\begin{document}

\title{Considerations Concerning the Radiative Corrections to Muon Decay in the Fermi and Standard Theories} 
\author{A.FERROGLIA, G.OSSOLA, and A.SIRLIN\footnote{Corresponding Author: A.Sirlin,\\ e.mail: alberto.sirlin@nyu.edu, tel:(212)998-7734, fax:(212)995-4016.}}
\address{Department of Physics, New York University,\\
  4 Washington Place, New York, NY 10003, USA}
\maketitle

\begin{abstract}
The FAC, PMS, and BLM optimization methods are applied to the QED corrections
to the muon lifetime in the Fermi V-A theory.
The FAC and PMS scales are close to $m_e$, while the BLM scale 
nearly concides with the geometric average $\sqrt{m_e m_\mu}$. 
The optimized expressions are employed to estimate the third
order coefficient in the $\alpha(m_\mu)$ expansion and the theoretical error 
of the perturbative series. Using arguments based on 
effective field theory and a simple examination of Feynman diagrams, 
it is shown that, if contributions of 
${\cal O}(\alpha m_\mu^2/M_W^2 )$
are neglected, the corrections to muon decay in the SM factorize
into the QED correction of the Fermi V-A theory and the electroweak amplitude 
$g^2/(1-\Delta r)$, both of which are strictly scale-independent.
We use the results to clarify how the QED corrections to muon decay
and the Fermi constant 
$G_F$ should be used in the SM, and what is the natural choice
of scales if running couplings are employed.
\end{abstract}

\newpage

\section{Introduction}

The one-loop radiative corrections to $\mu$-decay in the 
Fermi theory were  evaluated approximately four decades ago. 
These studies included the correction to the electron spectrum 
and muon lifetime $\tau_{\mu}$\cite{c1,c2}, as well as the 
momentum dependence of the asymmetry and the integrated asymmetry for polarized
muons\cite{c1}.
Although the Fermi theory is, in general, non-renormalizable, 
the $\mu$-decay case is especial.
For vector and axial vector interactions of the charge 
retention order, which are the relevant couplings in the two-component 
theory of the neutrino and in the V-A theory, a theorem assures us that, 
to first order in $G_{\mu}$, but all orders in $\alpha$, 
the corrections to $\mu$-decay are convergent after  mass and charge 
renormalization\cite{c3}.
A striking cancellation of mass singularities in the correction to the 
lifetime and integrated asymmetries was found, which occurs also for the
scalar, pseudoscalar, and tensor 
Fermi interactions, and the $\beta$-decay lifetime\cite{c1}. 
These observations were one of the motivations in the derivation of 
the KLN theorem\cite{c4,c5}.
Years later, in order to clarify a controversy that arose in the 
determination of the Fermi constant, the corrections of 
${\cal O}\left( \alpha^2 \ln{(m_\mu/m_e)} \right)$ to $\tau_{\mu}$
 were obtained\cite{c6}.

Very recently, in an important theoretical development, 
van Ritbergen and Stuart completed the evaluation of 
${\cal O}\left(\alpha^2 \right)$ corrections to $\tau_{\mu}$ in the 
local V-A theory, in the limit $m_{e}^2/m_{\mu}^2 \to 0$\cite{c7,c8}.
Their final answer can be expressed succintly as
\begin{equation} \label{eq1}
  C \left( m_{\mu} \right) = \frac{\ammu}{\pi} c_{1} 
              + \left(\frac{\ammu}{\pi}\right)^{2}c_2,
\end{equation} 
\begin{eqnarray} \label{eq2}
c_{1}=\frac{1}{2}\left(\frac{25}{4}- \pi^{2} \right) & ; & c_{2}=6.700\ . 
\end{eqnarray}
In Eq.(\ref{eq1}), $\ammu c_{1}/\pi$ is the one-loop 
result\cite{c1} and $\ammu$
is a running coupling defined by
\begin{equation} \label{eq3}
  \ammu = \frac{\alpha}{1- \left( \frac{2\alpha}{3\pi} 
       +\frac{\alpha^{2}}{2 \pi^{2}} \right) \ln{\frac{m_\mu}{m_e}}}.
\end{equation}
In Refs.\cite{c7,c8}, the contribution of the last term in the denominator of
 Eq.(\ref{eq3}) is separated out, as 
$\left( \alpha^{3}/2 \pi^{2} \right) \ln{(m_{\mu}/m_{e})}$. 
The two expressions differ by only $0.2$ ppm, so that we will employ the 
more succint expression of Eq.(\ref{eq3}).

In Section 2, we apply the traditional FAC\cite{c9}, PMS\cite{c10}, and 
BLM\cite{c11} optimization schemes to the expansion of Eq.(\ref{eq1}).
We use those methods for estimations of the third order coefficient 
and the theoretical error in Eq.(\ref{eq1}).
In Section 3 we turn our attention to the radiative corrections to muon 
decay in the Standard Model (SM). Using arguments based on 
effective field theory and simple considerations of Feynman diagrams, 
we show that, if terms of  ${\cal O}(\alpha m_\mu^2/M_W^2)$
are neglected, the radiative corrections factor out into $[1+C(m_e)]$
(defined in Section 2) 
and the electroweak amplitude $g^2/(1-\Delta r)$, which are separately 
scale-independent. 
We use the results to discuss the application of the Fermi constant 
$G_F$ in electroweak physics. Section 4 presents our conclusions. 

\section{Application of Optimization Methods}

As it is well known, $\alpha(\mu)$ does not run below the $m_e$ scale 
and $\alpha(m_e)$ is identified with the conventional fine structure constant, 
$\alpha^{-1}=137.03599959(38)(13)$\cite{c12}.
Setting then $\alpha=\alpha(m_e)$ in Eq.(\ref{eq3}) and replacing $m_e\to \mu$,
we have
\begin{equation} \label{eq4}
  \alpha(m_\mu)=\frac{\alpha(\mu)}{1-\left(\frac{2\alpha(\mu)}{3\pi}+
  \frac{\alpha^2(\mu)}{2\pi^2}\right)\ln{\frac{m_\mu}{\mu}}}.
\end{equation}
Inserting this expression into Eq.(\ref{eq1}), expanding and truncating
 in second order, we find 
\begin{equation} \label{eq5}
C(\mu)= \frac{\alpha(\mu)}{\pi}c_1+\left( \frac{\amu}{\pi} \right)^2
        \left[ c_2+\frac{2}{3}c_1\ln{\frac{m_\mu}{\mu}} \right].   
\end{equation}

The method of Fastest Apparent Convergence (FAC)\cite{c9} 
chooses the scale $\mu_\textrm{\tiny FAC}$
in such a manner that the NLO term vanishes. Thus, we have 
$ \ln{(m_\mu/\mu_\textrm{\tiny FAC})}=-3c_2/(2c_1)$,
which leads to $\mu_\textrm{\tiny FAC}=0.801\ m_e$. As $\amu$ 
does not run below $\mu=m_e$,
we interpret this result as $\mu_\textrm{\tiny FAC}=m_e$. 
Eq.(\ref{eq5}) becomes
\begin{equation} \label{eq7}
  C(\mu_\textrm{\tiny FAC})=C(m_e)=\frac{\alpha}{\pi}c_1
           +\left( \frac{\alpha}{\pi} \right)^2
        \left[ c_2+\frac{2}{3}c_1\ln{\frac{m_\mu}{m_e}} \right], 
\end{equation}
an expansion in terms of the fine structure constant $\alpha$.
The $\cal{O}$$\left( (\alpha/\pi)^2 ln{(m_\mu/m_e)}\right)$
term concides with the expression found in Ref.\cite{c6}. Recalling 
$m_\mu/m_e=206.768273(24)$\cite{c12}, we see that the two terms between square 
brackets nearly cancel each other and Eq.(\ref{eq7}) becomes
\begin{equation} \label{eq8}
  C(\mu_\textrm{\tiny FAC})=C(m_e)=\frac{\alpha}{\pi}c_1
             +\left( \frac{\alpha}{\pi} \right)^2 0.26724\ .
\end{equation}
 
The Principle of Minimal Sensitivity (PMS)\cite{c10} identifies 
$\mu_\textrm{\tiny PMS}$ 
with the
stationary point of $C(\mu)$. Applying $\mu (d/d\mu)$ to $C(\mu)$, 
and recalling the renormalization group equation (RGE)
\begin{equation} \label{eq9}
  \mu\frac{d}{d\mu}\amu=\frac{2}{3\pi}\alpha^2(\mu)+\frac{\alpha^3(\mu)}
                          {2\pi^2},
\end{equation}
we obtain
\begin{equation} \label{eq10}
  \mu\frac{d}{d\mu}C(\mu)=\frac{4}{3}\left( \frac{\amu}{\pi} \right)^3
                             \left[ c_2+\frac{3}{8}c_1+\frac{2}{3}c_1
                              \ln{\frac{m_\mu}{\mu}}  \right].
\end{equation}
Thus, the stationary point is given by
\begin{displaymath}
 \ln{\frac{\mu_\textrm{\tiny \tiny PMS}}{m_\mu}}=\frac{3}{2}\left(\frac{3}{8}+
                                  \frac{c_2}{c_1} \right),
\end{displaymath}
which leads to $\mu_\textrm{\tiny PMS}=1.40636\ m_e$. We note that Eq.(\ref{eq9}) is
consistent with
\begin{equation}   \label{eq11}
  \amu=\frac{\alpha}{1-\left( \frac{2\alpha}{3\pi}+\frac{\alpha^2}{2\pi^2}
              \right)\ln{\frac{\mu}{m_e}}},
\end{equation}
an expression that can be obtained by replacing $m_\mu\to \mu$, $\mu\to m_e$
in Eq.(\ref{eq4}), and will be useful in the following discussion.
Inserting $\mu=\mu_\textrm{\tiny PMS}$ in Eq.(\ref{eq5}), we obtain
\begin{equation} \label{eq12}
  C(\mu_\textrm{\tiny PMS})=\frac{\amupms c_1}{\pi}
     -\left( \frac{\amupms}{\pi} \right)^2
                  \frac{3}{8}c_1,
\end{equation}
or
\begin{equation} \label{eq13}
  C(\mu_\textrm{\tiny PMS})=\frac{\amupms c_1}{\pi}
      +\left( \frac{\amupms}{\pi} \right)^2 0.67868\ .
\end{equation}

The BLM method \cite{c11} chooses $\mu_\textrm{\tiny BLM}$ so as to 
cancel the terms in 
$c_2$ proportional to $n_f$, the number of light fermions. 
In the $\mu$-decay case, 
there is only one light fermion, the electron. From Ref.\cite{c7} we learn 
that the
electron loop contributions to $c_2$ (virtual loops and pair creation) is
$3.22034$. Splitting $c_2=3.22034+3.47966$, 
one chooses $\mu_\textrm{\tiny BLM}$ to cancel 
the first contribution in the second order coefficient of Eq.(\ref{eq5}).
Thus, $\ln{(m_\mu/\mu_\textrm{\tiny BLM})}=-(3/2) (3.22034/c_1)$,
which leads to $\mu_\textrm{\tiny BLM}=m_\mu/14.4267=14.3323\ m_e$. We note 
that this is very close to $\sqrt{m_e m_\mu}=14.38\ m_e$, the geometric 
average of the two masses. 
The BLM expansion is
\begin{equation} \label{eq14}
  C(\mu_\textrm{\tiny BLM})=\frac{\amublm c_1}{\pi}
           +\left( \frac{\amublm}{\pi} \right)^2 3.47966\ .
\end{equation}

The FAC, PMS, and BLM optimization expressions of Eqs.(\ref{eq8}, \ref{eq13}, 
\ref{eq14}), can be used to estimate the coefficient $c_{3}$ 
of $\left(\ammu/\pi \right) ^{3}$ in the expansion of Eq.(\ref{eq1}). 
It is well known that such estimations are not particularly 
useful if contributions of an important class of new diagrams open up 
at the relevant level.
For instance, in the radiative corrections to $g-2$, large contributions 
due to the light by light scattering open up already in 
${\cal O} \left( \alpha^{3} \right)$.
In $\mu$-decay we encounter a more felicitous situation, as light by light 
scattering contributes only in   $\cal{O}$$\left( \alpha^{4} \right)$. 
Writing the three optimized expansions in the generic form
\begin{equation}
  C \left( \mu^{*} \right) = \frac{\alpha (\mu^{*})}{\pi} c_{1} 
       + \left( \frac{\alpha (\mu^{*})}{\pi} \right) ^{2} c_{2}^{*},
\end{equation}
and expressing $\alpha (\mu^{*})$ in terms of $\ammu$ by replacing 
$m_{\mu} \to \mu^{*}$, $\mu \to m_{\mu}$ in Eq.(\ref{eq4}),
we find the estimation  
\begin{equation} \label{eq16}
\left(c_{3} \right)_{est} = \frac{4}{9} c_{1} 
    \ln^2{\frac{\mu^{*}}{m_{\mu}}} 
    + \left(\frac{c_{1}}{2} + \frac{4}{3} c_{2}^{*} \right) 
    \ln{\frac{\mu^{*}}{m_{\mu}}}.
\end{equation}
Using the values of $\mu^{*}$ and $c_{2}^{*}$ obtained above, 
Eq.(\ref{eq16}) leads to 
\begin{equation} \label{eq17}
\left( c_{3} \right)_{est} = \left\{ \begin{array}{ll}
-19.9 & \textrm{(FAC)} \\
-20.0 & \textrm{(PMS)} \\
-15.7 & \textrm{(BLM)}
\end{array} \right.
\end{equation}
We see that the three estimates are quite close.

It is also interesting  to compare the numerical results 
of the optimized expansions among themselves and with that of Eq.(\ref{eq1}).
Using the values of $\mu_\textrm{\tiny PMS}$ and $\mu_\textrm{\tiny BLM}$ 
found before, and evaluating
$\alpha(\mu^*)$ ($\mu^*=\mu_\textrm{\tiny PMS}$,$\mu_\textrm{\tiny BLM}$) 
via Eq.(\ref{eq11}),
the expansions of Eqs.(\ref{eq8},\ref{eq13},\ref{eq14},\ref{eq1}) lead to
\begin{eqnarray} \label{eq18a}
  C(m_e) & = &  -4.202402\times 10^{-3}\quad (FAC),\\
 C(\mu_\textrm{\tiny PMS}) & = &  -4.202403\times 10^{-3}\quad (PMS),\\ 
C(\mu_\textrm{\tiny BLM}) & = &  
         -4.202348\times 10^{-3}\quad (BLM),\\ \label{eq18d}
C(m_\mu) & = &  -4.202147\times 10^{-3}\quad  (\alpha(m_\mu)\ \textrm{exp.}).
\end{eqnarray}
We see that the three optimization methods give very close results, 
with a maximum absolute difference of $5.4\times 10^{-8}$. As
the important mass scales for $\mu$ decay in the Fermi theory are
$m_e$ and $m_\mu$, it is natural to consider $m_e<\mu<m_\mu$ as the relevant 
range. We may therefore take the maximum difference among the four 
evaluations above as an estimate of the theoretical error. 
This is given by the difference 
$C(m_\mu)-C(\mu_\textrm{\tiny PMS})=2.6\times 10^{-7}$, 
which is equivalent to the estimated third order contribution 
$20.0[\alpha(m_\mu)/\pi]^3$ in the PMS estimate [Cf.Eq.(\ref{eq17})].
This leads to an error of $1.3\times  10^{-7}$ in the determination of $G_F$, 
which is very close to the estimate given in Ref.\cite{c8} from the 
consideration of the known leading third order logarithm in $C(m_e)$.
We also recall that the one-loop corrections proportional to powers
and logarithms of $m_e^2/m_\mu^2$ can be obtained by combining Ref.\cite{c13}
and Ref.\cite{c1}, and have been reported in Refs.\cite{c8,c14}. When 
the tree-level $\mu$ decay phase space is factored out, the leading correction
 of this type is
 \begin{equation} \label{eq19}
   \frac{\alpha}{\pi}\frac{m_e^2}{m_\mu^2}\left[24log{\frac{m_\mu}{m_e}}
                    -9-4\pi^2 \right] = 4.3\times 10^{-6}.
 \end{equation}
In very high precision calculations, Eq.(\ref{eq19}) should be
added to Eqs.(\ref{eq18a}-\ref{eq18d}) 
so that the expansions, rounded to four decimals, become
\begin{eqnarray} \label{eq20s}
C(m_e)= C(\mu_\textrm{\tiny PMS}) & = &  -4.1981\times 10^{-3},\\ 
C(\mu_\textrm{\tiny BLM}) & = &  -4.1980\times 10^{-3},\\ 
C(m_\mu) & = &  -4.1978\times 10^{-3}.
\end{eqnarray}
The error due to the uncalculated terms of 
${\cal O}((\alpha/\pi)^2(m_e^2/m_\mu^2)\ln^2{m_\mu/m_e})$ 
has been estimated to be a few times $10^{-7}$\cite{c8}.
This seems quite resonable. In fact, we note that if the
${\cal O}((\alpha/\pi)^2(m_e^2/m_\mu^2))$ 
contributions had the same magnitude relative to the leading 
${\cal O}(\alpha^2)$ term $(\ammu/\pi)^2c_2$ in Eq.(\ref{eq1}), as the
${\cal O}((\alpha/\pi)(m_e^2/m_\mu^2))$ bear with respect to
$(\ammu/\pi)c_1$, their contribution would be even smaller, a few times 
$10^{-8}$. 

\section{Factorization of the Radiative Corrections to Muon Decay 
in the SM}

In the on-shell renormalization scheme of the SM, it is customary to write 
$1/\tau_\mu$  in the form
\begin{eqnarray} \label{venticinque}
  \frac{1}{\tau_\mu}& =& \frac{P}{32}\frac{g^4}{m_W^4}
                   \frac{[1+C(m_e)]}{(1-\Delta r)^2 },\\ \label{ventisei}
  P&=&f\left(\frac{m_e^2}{m_\mu^2}\right)\left[1+\frac{3}{5}
                     \frac{m_\mu^2}{M_W^2}\right]\frac{m_\mu^5}{192\pi^3},\\
 f(x)&=&1-8x-12x^2\ln{x}+8x^3-x^4, 
\end{eqnarray}
where $g^2=e^2/\sin^2{\theta_w}$, $\sin^2{\theta_w}=1-M_W^2/M_Z^2$,
$M_W$ and $M_Z$ are pole masses, 
$C(m_e)$ is the radiative correction of the Fermi V-A theory and $\Delta r$
is the electroweak radiative correction introduced in Ref.\cite{c15}.
Before the work of Refs.\cite{c7,c8}, only the terms involving $c_1$ in 
Eq.(\ref{eq7}) were known and $C(m_e)$ was approximated by the expression
\begin{displaymath}
  \frac{\alpha}{\pi}c_1\left[1+\frac{2\alpha}{3\pi}
           \ln{\frac{m_\mu}{m_e}}\right].
\end{displaymath}
In Eq.(\ref{ventisei}), $f(m_e^2/m_\mu^2)m_\mu^5/192\pi^3$ is a phase space factor
and $3m_\mu^2/5M_W^2$ is the tree-level contribution of the W-propagator.

If terms of ${\cal O}(\alpha m_\mu^2/M_W^2)$ are neglected in the radiative 
corrections, and $\Delta r $ is approximated by 
$\Delta r^{(1)}- e^4\textrm{Re}\Pi_2^{(r)}(M_Z^2)$, where $\Delta r^{(1)}$
is the one-loop contribution to $\Delta r$, and  $e^4\Pi_2^{(r)}(M_Z^2)$
is the two-loop contribution to the renormalized vacuum polarization function 
at $q^2=M_Z^2$, it was shown in Ref.\cite{c16} that Eq.(\ref{venticinque}) 
contains
all the one-loop corrections, as well as all two-loop contributions involving 
mass singularities (i.e. logarithms $\ln{m}$, where $m$ is a generic 
light fermion mass).
It also contains all the terms
of ${\cal O}\left[ \left( \alpha \ln(M_Z/m)\right)^n\right]$. In fact, 
if so desired, all these logarithms can be absorbed by expressing
$g^4/(1-\Delta r)^2$ in terms of the running coupling 
$\alpha(M_Z)= \alpha/(1-\Delta\alpha)$, where 
$\Delta\alpha=-\textrm{Re}\Pi^{(r)}(M_Z^2)$.
This result follows from the observation that, in 
$g^4/(1-\Delta r)^2$, such logarithms arise from the renormalization of 
$\alpha_0$ in terms of $\alpha$.

We now show that, if contributions of $\cal{O}$$(\alpha m_\mu^2/M_W^2)$ 
are neglected
in the radiative corrections, the factorization displayed in 
Eq.(\ref{venticinque}), 
involving the QED corrections of the Fermi theory and  the electroweak factor
$g^4/(1-\Delta r)^2$, is valid to all orders in perturbation theory.
This applies not only to the corrections to the lifetime, but also to those 
affecting the electron spectrum, as well as the momentum dependence of the 
asymmmetry and the integrated asymmetry in the case of polarized muons.
We present two arguments, one based on the effective field theory
approach\cite{c17}, the other involving a simple discussion of higher order
Feynman diagrams.

If contribution of
${\cal O}(\alpha m_\mu^2/M_W^2)$ are neglected in the radiative corrections
and the tree-level term $3m_\mu^2/5M_W^2$ in Eq.(\ref{ventisei}) is for the
moment disregarded, the effective field theory at the muon mass scale
is the local V-A four-fermion Lagrangian density
\begin{displaymath}
 {\cal L}\ =\ -\frac{ G_F}{\sqrt{2}}
 \left[ \overline{\Psi}_e\gamma^\mu(1-\gamma^5)\Psi_{\nu_e} \right]
 \left[\overline{\Psi}_{\nu_\mu}\gamma_\mu(1-\gamma^5)\Psi_{\mu} 
 \right],
\end{displaymath}
plus QED, plus QCD.
Therefore, one can systematically evaluate the corrections to the
spectrum, lifetime, and 
asymmetry in muon decay on the basis of this effective Lagrangian.
This procedure results in the usual expressions involving $G_F$ and
the radiative corrections of the Fermi V-A theory\cite{c1}. 
As the latter are 
convergent to all orders in $\alpha$\cite{c3}, there is no need to cancel 
ultraviolet divergences and, therefore, in their expression, there is no 
reference to the high mass scale $M_Z$ of the underlying theory.
As mentioned in the Introduction, these corrections are known at the one-loop
level in the case of the electron spectrum and asymmetry, and have now been
evaluated at the two-loop level in the case of $1/\tau_\mu$. In particular,
one finds
\begin{equation} \label{eq20}
  \frac{1}{\tau_\mu} = \frac{G_F^2 m_\mu^5}{192\pi^3}f(m_e^2/m_\mu^2) 
\left[1+C(m_e)\right],
\end{equation}
where $C(m_e)$ is the two-loop correction discussed in Sections 1 and 2. 
This leads to Eq.(\ref{venticinque}) provided we identify
\begin{equation} \label{eq21}
  \frac{G_F}{\sqrt{2}} = \frac{g^2}{8M_W^2}\frac{1+\frac{3}{10}
\frac{m_\mu^2}{M_W^2}}{(1-\Delta r)}\ . 
\end{equation}
Eq.(\ref{eq21}) has a very simple interpretation: it is the matching 
relation that expresses the coupling constant of the effective, low energy 
theory, in terms of the coupling constants and radiative corrections of the
underlying theory. It is very important to note that $C(m_e)$ and 
$g^2/(1-\Delta r)$ are separately scale-independent quantities.
The $\mu$-independence of $C(m_e)$ follows from the fact that, 
when expressed in terms of physical parameters such as 
$\alpha$, $m_e$, $m_\mu$, 
it is convergent to all orders of perturbation theory\cite{c3}.  
The $\mu$-independence of $g^2/(1-\Delta r)$ follows then from the fact that
it can be expressed in terms of physical observables via 
Eqs.(\ref{eq20},\ref{eq21}). As explained above, in the on-shell scheme of 
renormalization $g^2$ is defined in terms of $e^2$, $M_W$ and $M_Z$, which
are physical quantities and, therefore, $\mu$-independent.
It follows that $\Delta r$ is $\mu$-independent, an important property that
has been verified at the one-loop level\cite{c15}, and through
terms of ${\cal O}(g^4M_t^2/M_W^2)$ at the two-loop level\cite{c18}.

The same conclusion, concerning the factorization of QED and electroweak
corrections when terms of $\cal{O}$$(\alpha m_\mu^2/M_W^2)$ are neglected,
can be reached by a simple analysis of Feynman diagrams. Consider,
for instance, a two-loop diagram involving a photon of virtual momentum $k$
attached to external $\mu$ and/or $e$ lines and a second loop that 
includes heavy particles. The relevant momenta in the QED correction of
the Fermi theory are $|k^2|\lesssim m_\mu^2$. If we neglect terms
of $\cal{O}$$(\alpha m_\mu^2/M_W^2)$, we can set $k=0$ in the heavy-loop 
integration and the two-loops factor out. We also note that when $|k^2|$
becomes a few times larger than $m_\mu^2$, we can set $p_e=0$, where
$p_e$ is the electron four-momentum. All reference to $p_e$ is lost and 
we see that such domain of virtual momenta does not contribute
to the corrections to the electron spectrum. Thus, the latter involves 
essentially the domain  $|k^2|\lesssim m_\mu^2$, for which the
factorization is valid.
A second consideration, based on Feynman diagrams, refers to the factorization 
of QED corrections relative to the dominant electroweak corrections. These 
arise from the renormalization of the bare coupling constant 
$g_0^2=e_0^2/s_0^2$. Consider, for simplicity, the one-loop photonic diagrams
to $\mu$-decay in the SM. Each diagram carries a factor $e_0^2/s_0^2$ 
associated with the virtual $W$ interchange. 
The renormalization of $e_0^2/s_0^2$
in such diagrams leads, in higher orders, to dominant electroweak corrections
that clearly factorize with respect to the QED corrections. 
The same arguments can be extended to higher orders.

The crucial Eq.(\ref{eq21}) can be simplified by introducing 
a coupling constant
\begin{displaymath}
  G_\mu=\frac{G_F}{1+\frac{3m_\mu^2}{10M_W^2}}, 
\end{displaymath}
so that Eqs.(\ref{eq20},\ref{eq21}) become
\begin{eqnarray} \label{eq22}
\frac{1}{\tau_\mu} & =& \frac{G_\mu^2 m_\mu^5}{192\pi^3}f(m_e^2/m_\mu^2)
\left[1+\frac{3}{5}\frac{m_\mu^2}{M_W^2}\right] 
\left[1+C(m_e)\right],\\  \label{eq23}
  \frac{G_\mu}{\sqrt{2}} & = & \frac{g^2}{8M_W^2}\frac{1}{(1-\Delta r)}\ .
\end{eqnarray}
These are, in fact, the expressions most commonly used in literature. The
difference between $G_\mu$ and $G_F$ is $0.5$ ppm;  it is
completely negligible at present and it would only be of marginal 
interest if the experimental error in $\tau_\mu$ 
is reduced by a factor 10. Nonetheless, this factor
is frequently included in the theoretical expressions, 
as it is a tree-level contribution from the SM. Eq.(\ref{eq22}) has the 
convenient feature that all the kinematical factors of the SM
calculation are separated out.

The arguments carried above can be applied mutatis mutandi to other 
renormalization frameworks such as the $\overline{MS}$ scheme of 
Ref.\cite{c19},
in which couplings in $g^2/(1-\Delta r)$ are identified with running 
$\overline{MS}$ parameters evaluated at $M_Z$, masses are still interpreted 
as pole masses, and $\Delta r$ is modified accordingly.

The authors of Ref.\cite{c8} discuss a renormalization framework in which the 
couplings in  $g^2/(1-\Delta r)$ are also identified with running 
parameters evaluated at $M_Z$. In particular, they claim that, in order 
to ensure the consistency of the renormalization scheme, the QED correction
$C$ of the local theory, when applied to the SM, must be evaluated via 
Eq.(\ref{eq1}) with $\alpha(m_\mu)\to\alpha_e(M_Z)$ ($\alpha_e(M_Z)$ is
the expression obtained from Eq.(\ref{eq3}) by replacing $m_\mu\to M_Z$).
This claim has a curious consequence: before $c_2$ was evaluated, the current
experimental value of $\tau_\mu$ led, via Eq.(\ref{eq20}), to 
$G_F=1.16639(2)\times10^{-5}\ /GeV^2$. With the incorporation of the $c_2$ 
term in Eq.(\ref{eq1}) one obtains $G_F=1.16637(1)\times10^{-5}\ /GeV^2$.
However, according to the new claim, the correction $C$ is to be evaluated
differently when applied to the SM than in the Fermi theory, with
the result that the relevant value is turned back to
$G_F=1.16639(1)\times10^{-5}\ /GeV^2$!

In order to clarify this strange situation, we make the following 
observations:\\
i) As explained above, both the QED correction $C$ and the electroweak 
amplitude $g^2/(1-\Delta r)$ are separately $\mu$-independent quantities.
One can certainly evaluate these corrections using running couplings. In this 
case, if the expressions are carried out to all orders, the $\mu$-dependence 
of the couplings is cancelled by the $\mu$-dependence of the radiative 
corrections.
In actual calculations, due to the necessary truncation of the perturbative
series, a $\mu$-dependence emerges. However, there is no reason to require
that the same scale be employed in  $g^2/(1-\Delta r)$ and in $C(\mu)$.
Indeed, the natural scale in the former is $\mu\approx M_Z$ and, in the latter, it is somewhere in the range $m_e<\mu<m_\mu$, with the optimization methods 
discussed in Section 2 favoring the lower values of $\mu$.
For instance, part of $C$ are I.B. contributions. The photons here carry 
$k^2=0$ and it is clear that their natural coupling is $\ame=\alpha$, in sharp contrast with $\alpha_e(M_Z)$.\\
ii) Although the choice $\mu=M_Z$ for the QED correction is not natural, 
its implementation should be done by setting $\mu=M_Z$ in Eq.(\ref{eq5}),
rather than replacing $\ammu\to\alpha_e(M_Z)$ in Eq.(\ref{eq1}).
Using Eq.(\ref{eq5}) and $\alpha_e(M_Z)$ to evaluate $C(M_Z)$,
we find that $G_F$ would be 
decreased by only $0.6$ ppm rather than increased by $20$ ppm. In particular,
we note that changing the scale of $\amu$ in Eq.(\ref{eq1}), without modifying
the second order coefficient according to Eq.(\ref{eq5}), is inconsistent at 
the two-loop level.

In summary, our conclusion is that, subject to the neglect of terms of
$\cal{O}$$(\alpha m_\mu^2/M_W^2)$, one should apply the same QED corrections
in the SM as in the Fermi theory. In particular, the value 
$G_F=1.16637(1)\times10^{-5}\ /GeV^2$, obtained in the Fermi theory, is the 
one that should be applied in the analysis of the SM. 

\section{Conclusions}

In Section 2, we have applied the traditional FAC, PMS and BLM 
optimization methods to the radiative corrections to the muon lifetime
in the Fermi V-A theory. 
As the corrections are convergent, one expects on general
grounds the natural mass scale to be in the range $m_e<\mu<m_\mu$.
The analysis shows that the FAC and PMS approaches select a mass scale very 
close to $m_e$, while the BLM scheme leads to one very near the geometric
average $\sqrt{m_e m_\mu}$ of the two masses.
The FAC and PMS methods provide nearly identical estimations of the third order
coefficient in the $\ammu$ expansion, while the BLM estimation is of the 
same sign and differs only by $20\%$ in magnitude. The three optimized 
expansions give very close results, with a maximum absolute difference
of $5.4\times 10^{-8}$. We use the maximal difference between the optimized 
and $\ammu$ expansion, which amounts to be $2.6\times 10^{-7}$, as an estimate
of the theoretical error due to the truncation of the perturbative series.
This translates into an error of $1.3\times 10^{-7}$ in the determination
of $G_F$, a result very close to estimates obtained in Ref.\cite{c8} by
different considerations.
We also find that the expansion in powers of the conventional fine
structure constant $\alpha$, which in this case is essentially equivalent
to the FAC expansion, contains a very small second-order coefficient.
In fact, the second-order term is about $2900$ times smaller
than the first. Thus, it turns out that, by a curious cancellation of two-loop
effects, the original one-loop calculation\cite{c1}, expressed in terms of $\alpha$,
had an error of only $1.4\times 10^{-6}$! Of course, this amusing 
historical fact could not be known before the work of Refs.\cite{c7,c8}
was carried out.

In Section 3, we return to the radiative corrections to muon decay in the 
SM. Using arguments based on the effective field theory approach, 
as well as considerations involving higher order Feynman diagrams, we show 
that, subject to the neglect of terms of  ${\cal O}(\alpha m_\mu^2/M_W^2)$,
the overall answer factorizes into two separately scale-independent 
corrections: those of the Fermi V-A theory and the electroweak amplitude
$g^2/(1-\Delta r)$. This important property applies to the corrections to
the various observables, such as the electron spectrum, electron asymmetry 
in the case of polarized muons, and the muon lifetime.
Therefore, if running couplings are employed, the scales may be chosen 
judiciously and independently in both factors, with $\mu\approx M_Z$ 
being the natural scale in  $g^2/(1-\Delta r)$, and $m_e<\mu<m_\mu$
being the logical range in the corrections of the Fermi V-A theory.
We reach the conclusion that, subject to the neglect of terms of 
 ${\cal O}(\alpha m_\mu^2/M_W^2)$, one should apply the same QED 
corrections to muon decay in the SM as in the Fermi theory. In particular, 
at variance with a claim  presented in Ref.\cite{c8}, we find that the value
$G_F=1.16637(1)\times 10^{-5}/GeV^2$, currently obtained in the Fermi theory,
is the one that should be applied in the analysis of the SM.

We conclude our discussion with the following comments:\\
i) Although the shifts discussed in Refs.\cite{c7,c8} and this paper 
are numerically very small, it is clearly desiderable to evaluate 
fundamental parameters such as $G_F$, as accurately as possible.\\
ii) At present, the Fermi V-A theory is generally viewed, not as independent, 
but rather as an effective low-energy theory derived from the SM.
The arguments presented in this paper make clear that the same constant $G_F$
that is precisely determined in the low-energy theory is, 
to a high degree of accuracy, the relevant paramenter in the analysis 
of the more fundamental, underlying gauge theory. 

\section*{Acknowledgements}

One of us (A.S.) would like to thank L.Dixon and M.Schaden for very
useful observations. This research was supported in part 
by NSF Grant No. PHY-9722083.


\begin{thebibliography}{99}
 \bibitem{c1}
           T.Kinoshita and A.Sirlin, Phys.Rev.\textbf{113} 
             (1959) 1652.     
 \bibitem{c2}
           S.M.Berman, Phys.Rev.\textbf{112}
             (1958) 267.   
 \bibitem{c3}
           S.M.Berman and A.Sirlin, Ann.Phys.\textbf{20} 
             (1962) 20. 
 \bibitem{c4}
           T.Kinoshita, J.Math.Phys.\textbf{3} 
             (1962) 650. 
 \bibitem{c5}
           T.D.Lee and M.Nauenberg, Phys.Rev.\textbf{133} 
             (1964) 1549. 
 \bibitem{c6}
           M.Roos and A.Sirlin, Nucl.Phys.\textbf{B29} 
             (1971) 296. 
 \bibitem{c7}
           T. van Ritbergen and R.G.Stuart, Phys.Rev.Lett.\textbf{82} 
             (1999) 488. 
 \bibitem{c8}
           T. van Ritbergen and R.G.Stuart, hep-ph/9904240.
 \bibitem{c9}
           G.Grunberg, Phys.Lett.\textbf{B95} (1980) 70 and 
           \textbf{B110} (1982) 501;\\ Phys.Rev.\textbf{D29} (1984) 2315. 
 \bibitem{c10}  
           P.M.Stevenson, Phys.Lett.\textbf{B100} (1981) 61; 
           Phys.Rev.\textbf{D23} (1981) 2916;\\
           Nucl.Phys.\textbf{B203} (1982) 472 and \textbf{B231} (1984) 65.
 \bibitem{c11}
           S.J.Brodsky, G.P.Lepage, and P.B.Mackenzie, Phys.Rev.\textbf{D28} 
           (1983) 228;\\ S.J.Brodsky and H.J.Lu,  
           Phys.Rev.\textbf{D51} (1995) 3652.
 \bibitem{c12}
           A.Czarnecki and W.J.Marciano, hep-ph/9810512.
 \bibitem{c13}
           R.E.Behrends, R.J.Finkelstein, and A.Sirlin,
           Phys.Rev.\textbf{101} (1956) 866.
 \bibitem{c14}
           Y.Nir, Phys.Lett.\textbf{B221} (1989) 184.
 \bibitem{c15} 
           A.Sirlin, Phys.Rev.\textbf{D22} (1980) 971.
 \bibitem{c16} 
           A.Sirlin, Phys.Rev.\textbf{D29} (1984) 89.
 \bibitem{c17}
           For a recent, illuminating discussion, see S.Weinberg, \emph{The 
           Quantum Theory of Fields}, (Cambridge Univ. Press, 1995), 
           in particular Section 12-3.
 \bibitem{c18} 
           G.Degrassi, P.Gambino, and A.Sirlin, 
           Phys.Lett.\textbf{B394} (1997) 188.
 \bibitem{c19} 
           A.Sirlin, Phys.Lett.\textbf{B232} (1989) 123;
           S.Fanchiotti and A.Sirlin, Phys.Rev.\textbf{D41} (1990) 319;
           G.Degrassi, S.Fanchiotti, and A.Sirlin, Nucl.Phys.\textbf{B351}
           (1991) 49.
\end{thebibliography}
\end{document}